\documentclass[eqsecnum,aps]{revtex4}
\newcommand{\beq}{\begin{equation}}
\newcommand{\eeq}{\end{equation}}

\begin{document}

\title{How Many Templates for GW Chirp Detection? The Minimal-Match Issue Revisited}
\author{R. P. Croce,  Th. Demma, V. Pierro, and I. M. Pinto}
\address{Wavesgroup, University of Sannio at Benevento, Italy}
\author{M. Longo, S. Marano and V. Matta}
\address{D.I.$^{3}$E., University of Salerno, Italy}

\begin{abstract}
In a recent paper dealing with maximum likelihood detection
of gravitational wave chirps from coalescing binaries with unknown parameters
we introduced an accurate representation
of the no-signal cumulative distribution of the  supremum
of the whole correlator bank.
This result can be used to derive a refined estimate 
of the number of templates yielding
the best tradeoff between detector's performance 
(in terms of lost signals among those potentially detectable)
and computational burden.
\end{abstract}
\maketitle

\section{Introduction}

Optimizing the design of the correlator-bank 
for detecting gravitational wave (henceforth GW) chirps 
from coalescing binaries (henceforth CB)
with unknown parameters is a key and yet not settled problem.
The main goal is to identify and adopt the best tradeoff 
between detector's performance and computational cost.
In a recent paper \cite{false_alarm}
we introduced  an accurate representation
of the cumulative distribution function (henceforth CDF) 
of the supremum of a bank of correlators in the null hypothesis
taking into account the nonzero covariance among them,
extending previous results by Mohanty \cite{Mohanty}. 
These findings can be used to re-address the following questions: 
i) whether it really makes sense to use 
the {\em largest} number of correlators
compatible with the available computing power, 
and
ii) whether the simple argument \cite{Apostolatos} whereby the 
{\em fraction} of potentially observable sources which would be lost
as an effect of template sparsity is ($1-\Gamma^3$),
$\Gamma$ being the minimal match \cite{Owen}
is still valid when one takes into account 
{\em both} the dependence of the detection threshold 
on the total number of correlators, 
{\em and} the covariance among the correlators,
at a given false-alarm level.
The above issues are discussed in Sect. 3 and 4 below, 
while Sect. 2 is aimed at introducing the relevant notation.

\section{Background}

The maximum likelihood (henceforth ML) strategy
for detecting GW chirps from CBs
with unknown parameters 
goes through the following steps.
First, a set of $N_\Delta$ reduced templates\footnote{
Reduced chirp templates are obtained by setting the template
coalescing time  and coalescing phase both equal to zero.}
is set up, viz.:
\beq
T(f,\vec{\xi}_T)=T_0 f^{-7/6}\exp[j\psi(f,\vec{\xi}_T)],
\label{eq:fraccazzill4}
\eeq
where $\psi(f,\vec{\xi}_T)$ is the (post-newtonian) chirp phasing
function \cite{OweSat}, $\vec{\xi}_T$ represents the {\em intrinsic}
source parameters (binary companion masses and spin parameters),
and $T_0$ is a suitable constant.
Second, a set of non-coherent reduced correlators
\beq
C[A,T]=
\sup_{\tau}
\left|
{\cal F}_{f \rightarrow \tau}
K_A[f,\vec{\xi}_T]
\right|
\label{eq:fraccazzill1}
\eeq
is computed, where ${\cal F}$ is the Fourier-transform operator and:
\beq
K_A[f,\vec{\xi}_T]=
\left\{
\begin{array}{l}
\displaystyle{
2~\frac
{
A(f)T^*(f,\vec{\xi}_T)
}
{\Pi(f)}
},~~~f \in (f_i,f_s)
\\
0,~~~~~~~~~\mbox{elsewhere}.
\end{array}
\right.
\label{eq:fraccazzill3}
\eeq
In eq. (\ref{eq:fraccazzill3}) $A(f)=S(f)+N(f)$ 
are the (spectral, noisy) data,
resulting from the superposition 
of a (possibly null) signal $S(f)$ 
and a realization $N(f)$ of the antenna noise, 
$(f_i,f_s)$ is the antenna spectral window,
and $\Pi(f)$ is the (one-sided) 
antenna noise power spectral density (henceforth PSD).
The Fourier transform in (\ref{eq:fraccazzill1})
is implemented using a discrete finite algorithm (DFT).
The corresponding number of coalescency times will be
henceforth denoted by $N_\Theta$; 
each reduced template is thus used
to compute $N_\Theta$ (non-reduced) correlators.\\
Once all correlators have been evaluated,
the final step consists in comparing the supremum 
\beq
\Lambda=
\sup_{\vec{\xi}_T}
C[A,T]
\label{eq:fraccazzill}
\eeq
(or any nondecreasing function thereof) 
to a threshold $\gamma$ depending 
on the prescribed false-alarm probability
\beq
\alpha=\mbox{prob}[\Lambda > \gamma|\mbox{no signal}].
\eeq 
The threshold $\gamma$ can be accordingly computed 
from the no-signal cumulative distribution of $\Lambda$.
Whenever $\Lambda > \gamma$ one declares 
that a signal has been observed 
(within false alarm probability $\alpha$), 
and estimates the unknown source parameters 
from those of the template 
yielding the largest correlator.\\
On the other hand, it can be shown that the false-dismissal probability 
\beq
\beta=\mbox{prob}[\Lambda \leq \gamma|\mbox{nonzero signal}]
\eeq
is a decreasing function of the match 
\beq
M[S,T]=
\left(
\int_{f_i}^{f_s}
f^{-7/3}\frac{df}{\Pi(f)}
\right)^{-1}
\sup_{\tau}
\left|
{\cal F}_{f \rightarrow \tau}
K_S[f,\vec{\xi}_T]
\right|,
\eeq
between the signal $S$ and the 
(most similar available) 
template $T$, where:
\beq
K_S[f,\vec{\xi}_T]=
\left\{
\begin{array}{l}
\displaystyle{
\frac{
f^{-7/3}\exp\left\{
j\left[
\psi(f,\vec{\xi}_S)-\psi(f,\vec{\xi}_T)
\right]
\right\}
}
{\Pi(f)}},~~~f \in (f_i,f_s)\\
0,~~~~~~~~~\mbox{elsewhere}.
\end{array}
\right.
\label{eq:fraccazzill6}
\eeq
The match is a measure of the similarity between $S$ and $T$, 
and is obviously equal to one iff  $S=T$.
The reduced-template set should be accordingly designed 
so as to cope with the more or less obvious prescription
\beq
\forall S \in {\cal S},
\exists T \in {\cal L}~:~
M[S,T] \geq \Gamma,
\eeq
where ${\cal S}$ and ${\cal L}$ are the allowed signal 
and template sets, respectively, 
and $\Gamma$ is the {\em minimal match} \cite{Owen}.

\section{Minimal Match vs. Number of Templates}

The natural question is 
how many reduced templates one should use in practice.
The standard argument suggests using  
the largest total {\em number} $N_{\Delta}$ of (reduced) templates
in the allowed (reduced) parameter-space search subset,
{\em compatible} with the available computing power,
so as to achieve the largest-minimal-match $\Gamma$
over the class of admissible sources.\\
In this connection, it should be stressed first that 
a visible knee point  
in the $\Gamma$ vs. $N_\Delta$
curve invariably exists 
at some value $N_\Delta=N_*$,
beyond which {\em any} further increase in $N_\Delta$ yields only
a {\em modest} increase in $\Gamma$.\\
This is illustrated in Fig.1
for the simplest case of the one-parameter family 
of (reduced) newtonian waveforms
in the range of chirp masses $(0.2M_\odot \leq {\cal M} \leq 10 M_\odot)$.
To compute the curve in Fig. 1 
we used the (initial) LIGO noise PSD \cite{LIGO_noise},
\beq
\Pi(f)=\frac{1}{4}
\left\{
1+\left(\frac{f_0}{f}\right)^4+
2\left(\frac{f}{f_0}\right)^2
\right\},
\label{eq:ligonoise}
\eeq
with $f_0=300$ $Hz$,  $f_i=40Hz$ and $f_s=400Hz$;
we sampled the data at $3.2KHz$ ($4$-times the Nyquist rate),
and let $N_\Theta=2^{21}$, which corresponds
to the longest-lived source allowed (${\cal M}=0.2 M_{\odot}$).\\
A knee point separating intervals 
of different (almost constant) slope
in the $\Gamma$ versus $N_\Delta$ curve
occurs in Fig. 1  at $N_\Delta \approx 5\cdot 10^3$.
Except for the numerical value of $N_*$, 
the existence of the above knee point 
is a {\em generic} feature of the $\Gamma$ vs. $N_{\Delta}$ curve, 
occurring also for more realistic, higher-order post-newtonian 
waveforms and templates.

\section{Undetected Observable Sources vs. Number of Templates}

It was first argued in \cite{Apostolatos} that
(under the simplest assumption of isotropically 
and homogeneously distributed sources)
the {\em fraction} of potentially observable sources 
which would be undetected as an effect of template mismatch is 
$(1-\Gamma^3)$.\\
This argument conceals an obvious naivety insofar
as it neglects the dependence of the detection threshold $\gamma$ 
corresponding to a given false-alarm probability $\alpha$
on the total number of templates $N=N_\Theta N_\Delta$,
and hence on $\Gamma$ itself.\\
In order to gain insight into this issue, it is expedient
to work in terms of the detection probability $P_d$, 
and capitalize on the accurate approximate representations
for the no-signal cumulative distribution of the whole-bank supremum
obtained in \cite{false_alarm}, which is needed to compute
the threshold $\gamma$ corresponding to a given $\alpha$.\\
Accordingly, let 
\beq
P_d=1-\beta=1-Q[\gamma-E(\Lambda)]
\label{eq:det_prob}
\eeq
where $Q(\cdot)$ is an unknown CDF, whose density can be nonetheless
safely assumed as being unimodal and center-symmetric, at least
for meaningful values of the signal-to-noise ratio.
In the worst case of {\em minimal} match,
\beq
E(\Lambda) \approx \Gamma d,
\eeq
where $d$ is the signal deflection (signal-to-noise ratio), viz.:
\beq
d:=\left|2 
\int_{f_i}^{f_s}
\frac{S(f)S^*(f)}{\Pi(f)} df
\right|^{1/2},
\label{eq:deflec}
\eeq
and we assume $d \gg 1$.
Let the {\em observable} sources be those for which $P_d \geq 0.5$.             
Under the above assumptions for $Q(\cdot)$, these correspond to:
\beq
d \geq \frac{\gamma}{\Gamma}.
\label{eq:defletto}
\eeq 
The deflection $d$ on the other hand can be written
\beq
d=\frac{K}{R}
\eeq
where $R$ is the source distance, and $K$ is a constant 
depending on the antenna and source  features, as well
as their mutual orientation.
Equation (\ref{eq:defletto}) allows to define 
the radius $R$ of the antenna-centered sphere
which contains all sources of a given class
(seen by the same antenna, sharing the same features and orientation) 
producing a deflection $d\ge\gamma$ (i.e., {\em observable}), viz.:
\beq
R=\frac{K\Gamma}{\gamma}.
\eeq
Hence the number of {\em observable} sources
in the given class, 
under the most obvious hypothesis of uniform 
(homogeneous and isotropic) spatial source density $\rho_s$ is:
\beq
N_s=\frac{4}{3}\pi\rho_s
\left[
\frac{K\Gamma}{\gamma}
\right]^3.
\label{eq:no_of_sources}
\eeq
Letting $\Gamma_{max}$, $N_{max}$ the largest achievable
minimal match, and the related number of templates,
and denoting as $N_s^{(max)}$ the corresponding value of $N_s$,
one has from (\ref{eq:no_of_sources}):
\beq
\eta=\frac{N_s}{N_s^{(max)}}=
\left(\frac{\Gamma}{\Gamma_{max}}\right)^3
\left[\frac{\gamma(\alpha,N_{max})}{\gamma(\alpha,N)}\right]^3
\label{eq:eta}
\eeq
where the dependence of $\gamma$ from $\alpha$ and $N$ has
been written explicitly. The quantity $1-\eta$ represents 
the fraction of {\em observable} sources which would
be lost due to poor minimal match ($\Gamma < \Gamma_{max}$).
Note that the ratio $\eta$ does {\em not} depend on $K$.\\
Neglecting the dependance of $\gamma$ on $N$ (and hence on $\Gamma$),
the second factor on the r.h.s. of (\ref{eq:eta}) cancels, 
and we are left with
\beq
\eta=\left(\frac{\Gamma}{\Gamma_{max}}\right)^3:=\eta_0,
\label{eq:eta0}
\eeq
which does {\em not} depend on $\alpha$.
Under the same assumption, $N_s$ in (\ref{eq:no_of_sources}) 
increases monotically with  $\Gamma$, 
attaining its supremum $N_s^{(max)}$ at $\Gamma_{max}=1$,
yielding $\eta_0=\Gamma^3$.
This is, in essence, the simplest argument 
introduced in \cite{Apostolatos}.\\
In order to refine this reasoning one should obviously include the
dependence of $\gamma$ on $N$, which in turn is affected by
the covariance among the correlators \cite{false_alarm}.
For illustrative purposes, we shall again stick 
at the simplest case of newtonian signals and templates, 
using the same parameters as in Fig. 1, 
and use the {\em best} available approximant
for the no-signal CDF of the detection statistic $\Lambda$ 
from \cite{false_alarm} in order to compute the ratio
$\gamma(\alpha,N_{max})/\gamma(\alpha,N)$ on the r.h.s. of 
(\ref{eq:eta})\footnote{
Note that $\gamma(\alpha,N_{max})/\gamma(\alpha,N) \geq 1$,
since $\gamma(\alpha,N)$ is a non-decreasing function of $N$,
whatever $\alpha$.}.
In Fig. 2  we accordingly compare $\eta_0$ (eq. (\ref{eq:eta0}), dashed line),
to $\eta$ (eq. (\ref{eq:eta}), solid lines), 
for $\alpha=10^{-k},~k=2,3,4$ (top-to bottom),
as functions of $N_\Delta$.
In Fig. 2 we assumed $\Gamma_{max}=0.99$ as a {\em bona fide}
practical limit value, and computed $N_{max}$ accordingly.
From Fig. 2 it is seen that a knee-point
in the curves exists at $N_\Delta \approx N_*\approx 5\cdot 10^3$. 
It is also seen that $\eta$ exceeds by a non-negligible $5\%$ typ. 
the plain estimate $\eta_0$,
for $N_\Delta \approx N_*$, if the best available approximant 
for the no-signal CDF of the detection statistic $\Lambda$
is used to compute the threshold\footnote{
Note that the referred approximant 
is still a lower bound for the true (unknown)
cumulative distribution \cite{false_alarm}, 
which accordingly provides 
a {\em conservative} estimate of the threshold.}.\\
It is also interesting to compare in terms of $\eta$
alternative (less accurate, i.e., 
yielding larger thresholds for the same $\alpha$ and $N$) 
approximants for the no-signal CDF of the detection statistic $\Lambda$.
This is obtained  by computing $N_s^{(max)}$ using the
{\em best} available approximant, i.e., using 
this latter to compute $\gamma(\alpha,N_{max})$ in (\ref{eq:eta}).
The resulting curves are shown in Fig. 3, where again 
we assumed $\Gamma_{max}=0.99$ as a {\em bona fide}
practical limit-value, and computed $N_{max}$ accordingly.
Specifically, the (scaled) quantities $\eta^{(0,1,2)}$ in Fig. 3
correspond to ignoring  the covariance among the correlators (dotted line), 
and to including the covariance among nearest-neighbour correlators
along one (coalescency-time, dashed line)
or both (coalescency-time and chirp-mass, solid line, best available approximant) 
coordinates of the newtonian parameter-space, respectively \cite{false_alarm}.
For each case, the $\eta$ vs. $N_\Delta$ curves corresponding to
$\alpha=10^{-2},10^{-3},10^{-4}$ are displayed. 
It is seen that the dependence of $\eta$ on $\alpha$ becomes
the less relevant, the better approximant is used.\\

\section{Conclusions}

On the basis of the above findings, the following conclusions can be drawn.
Inspection of Fig.s 1 and 2, shows a consistent evidence 
that increasing the number of templates
beyond a critical value corresponding to the knee-point 
in the curve of $\Gamma$ vs. $N_\Delta$ (Fig. 1),
does {\em not} produce a sensible increase in the detectable fraction
of potentially observable sources, at the expense of a marked growth
of computational load.
On the other hand, using the presently most accurate available
representation of the no-signal cumulative distribution of the
(whole-bank) detection statistic to compute the threshold
results in a sizeable increase ($\geq 5\%$)
in the detectable fraction of potentially
observable sources over the naive $\propto \Gamma^3$ estimate.\\
Both findings above support the current trend \cite{hierarchical}
toward the use of {\em hierarchical} search strategies 
for best tradeoff between detector's performance and computational burden.

\section*{Acknowledgements}

This work has been sponsored in part 
by the European Community through a Senior Visiting Scientist Grant 
to I.M. Pinto at NAO - Spacetime Astronomy Division, Tokyo, Japan, 
in connection with the TAMA project. 
I.M. Pinto wishes to thank all the TAMA staff at NAO, 
and in particular prof. Fujimoto Masa-Katsu 
and prof. Kawamura Seiji for kind hospitality 
and stimulating discussions.

\section*{References}


\newpage

\section*{Captions to the Figures}

$$~$$
Figure 1 - Minimal match vs. number of reduced templates
in $0.2M_{\odot} \leq {\cal M} \leq 10M_{\odot}$. Newtonian waveforms, Ligo-I noise.
$$~$$
Figure 2 - Detectable fraction of potentially observable sources,
in $0.2M_{\odot} \leq {\cal M} \leq 10M_{\odot}$.
$\eta_0$ (eq. (\ref{eq:eta0}), dashed line) and $\eta$ (eq. (\ref{eq:eta}), solid lines)
vs. $N_\Delta$, for $\alpha=10^{-k},~k=2,3,4$ (top-to bottom).
Newtonian waveforms, Ligo-I noise.
$$~$$
Figure 3 - Detectable fraction of potentially observable sources,
with $0.2M_{\odot} \leq {\cal M} \leq 10M_{\odot}$
vs. number of reduced templates at different false-alarm levels, using
different models of the no-signal cumulative distribution of the detection statistic.
Newtonian waveforms, Ligo-I noise.


\begin{thebibliography}{99}
\bibitem{false_alarm}{R.P. Croce et al., Class. Quantum Grav. {\bf 20} (2003) S803.}
\bibitem{Mohanty}{S.D. Mohanty, Phys Rev. {\bf D57} (1998) 630.}
\bibitem{Apostolatos}{Apostolatos Th. Phys. Rev. {\bf D54} (1996) 2421.}
\bibitem{Owen}{B.J. Owen  Phys. Rev. {\bf D53} (1996) 6749.}
\bibitem{OweSat}{B. Owen and B.S. Sathyaprakash, Phys. Rev. {\bf D60} (1999) 022002.}
\bibitem{LIGO_noise}{L.S. Finn and D. Chernoff, Phys. Rev. {\bf D47} (1993) 2198.}
\bibitem{hierarchical}{S. D. Mohanty and S. V. Dhurandhar, Phys. Rev. {\bf D 54} (1996) 7108.}
\end{thebibliography}
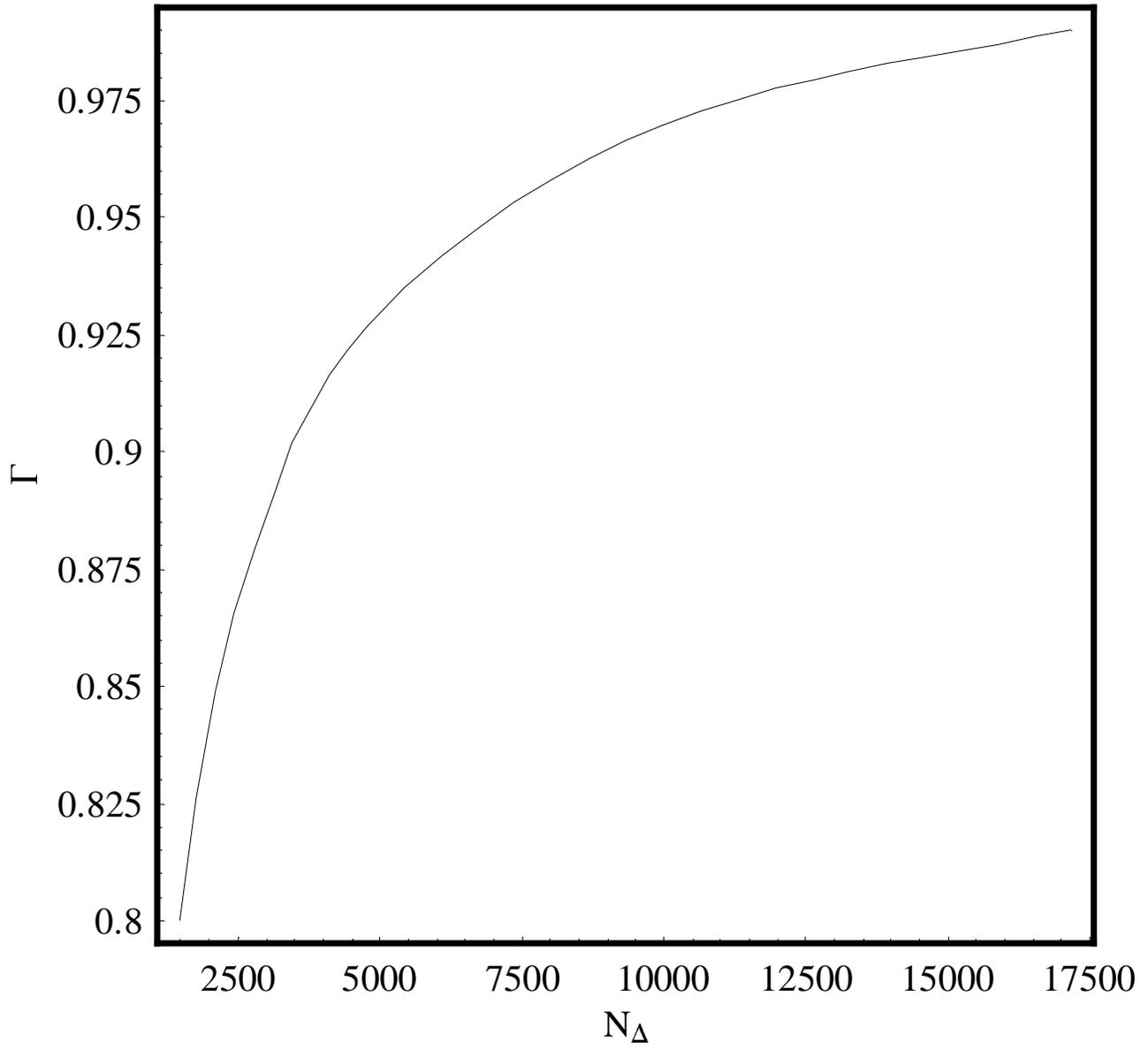
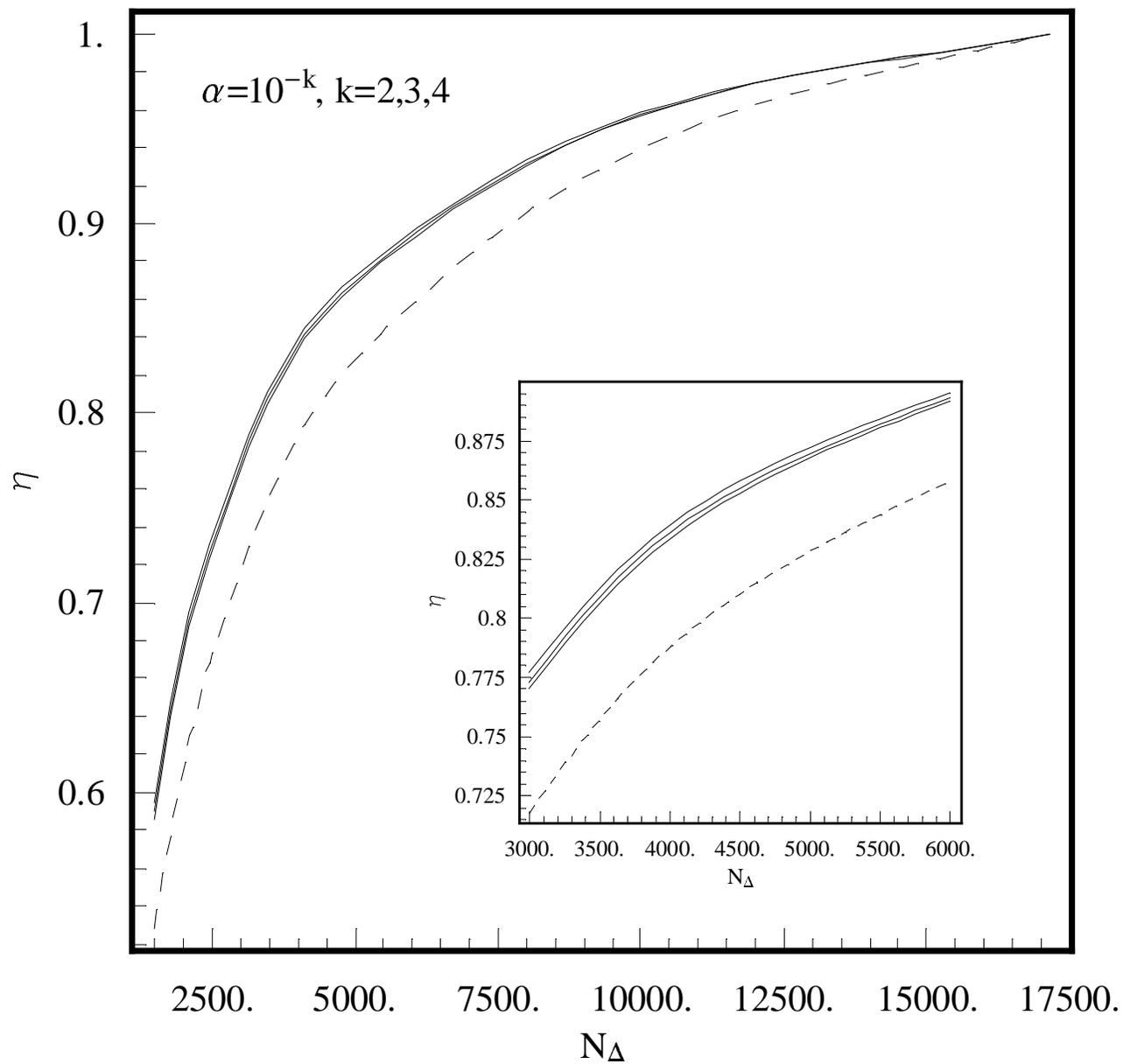
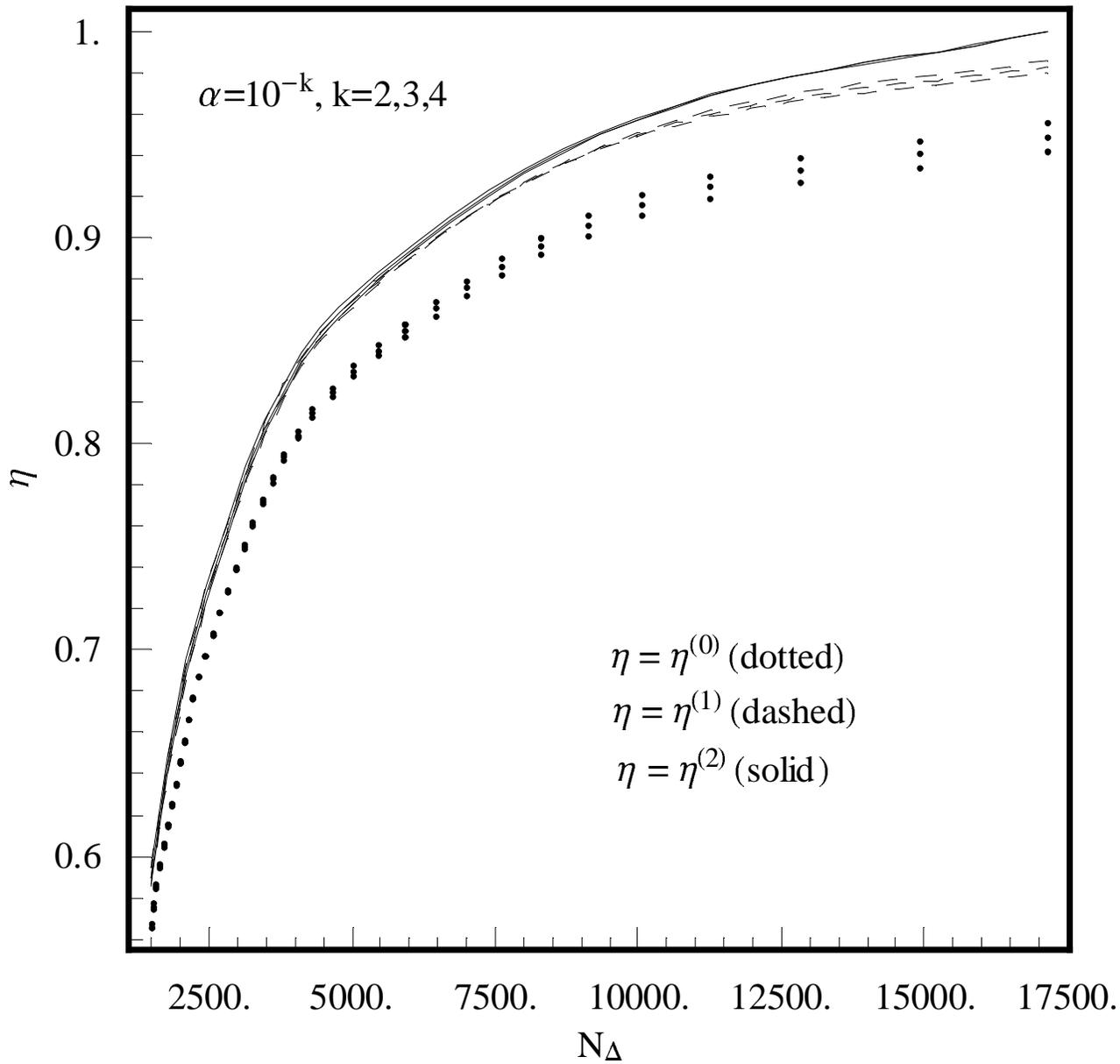
\end{document}